\documentstyle[referee,epsf]{MN}

\title[A New Statistical Indicator of Nonlinear Gravitational Clustering]
      {A New Statistical Indicator to Study Nonlinear \\
	Gravitational Clustering and Structure Formation}
\author[J.S.Bagla and T.Padmanabhan]
       {J.S.Bagla\thanks{E-mail : jasjeet@iucaa.ernet.in} and
        T.Padmanabhan\thanks{paddy@iucaa.ernet.in}\\
	Inter-University Centre for Astronomy and Astrophysics, Post Bag 4,
        Ganeshkhind, Pune 411 007, INDIA}

\date{Submitted to MNRAS}

\pagerange{\pageref{firstpage}--\pageref{lastpage}}
\pubyear{1995}

\begin{document}
\label{firstpage}

\maketitle

\begin{abstract}
In  an $\Omega=1$ universe dominated by nonrelativistic matter,
velocity field and gravitational force field are proportional to
each other in the linear regime. Neither of these quantities evolve in
time and these can be scaled suitably so that the constant of
proportionality is unity and velocity and force field are equal. The
Zeldovich approximation extends this
feature beyond the linear regime, until formation of pancakes.
Nonlinear clustering which takes place {\it after} the breakdown of
Zeldovich approximation, breaks this relation and the mismatch between
these two vectors increases as the evolution proceeds. We suggest that
the difference of these two vectors could form the basis for a
powerful, new,  statistical indicator of nonlinear clustering. We
define an indicator called velocity contrast, study its behaviour using
N-Body simulations and show that it can be used effectively to
delineate the regions where  nonlinear clustering has taken place. We
discuss several features of this statistical indicator and provide
simple analytic models to understand its behaviour. Particles with
velocity contrast higher than a threshold have a correlation function
which is biased with respect to the original sample. This bias factor
is scale dependent and tends to unity at large scales.
\end{abstract}


\section{Introduction}

Large scale structures like galaxies etc. are believed to have formed
out of small density perturbations via gravitational instability. This
process, in most popular models, is driven by dark matter which is the
dominant constituent of the universe. We can compute rate of growth of
clustering usint linear theory when the perturbations are
small.. Linear theory however has a very limited domain of validity
and we have to resort to numerical simulations for studying evolution
of inhomogeneities at late epochs.

In linear regime density field is related to velocity field in a
unique manner [ in the growing mode] and the density field alone
specifies the system completely. Evolution of perturbations is
described by a second order differential equation and specification of
initial density field and velocity field completely determine the
state of the system at any later time. However for a given nonlinear
density field there is no practical method for computing velocity field.
Our understanding of nonlinear regime will improve if we have a
simple physical indicator of velocity field. We introduce velocity
contrast, a new statistical indicator, that may be used to quantify
some features of velocity field.

Velocity contrast can be used for comparison of simulation results
with observations. It provides a simple and stable algorithm in
contrast with some other methods that are used for this
purpose. These methods are required as studies of nonlinear
gravitational clustering have focussed mainly on
aspects relating to dark matter. Comparison of these studies with
observations is made difficult by the fact that we observe only
sources of light. Many techniques have been devised for isolating
regions that can host galaxies in numerical simulations. Some of these
are quite elaborate, like DENMAX {\cite{denmax}}, and hence
computationally very intensive. Other schemes, like density threshold
or the friend of friend algorithm are very simple to
implement but have problems with interlopers as these algorithms do
not use dynamical information. We show that velocity contrast can be
used to isolate regions of interest in a relatively simple and robust
manner.

In next section, we briefly review dynamical evolution of
trajectories in a system undergoing gravitational collapse. This is
used to motivate the form of new indicator which is introduced in
\S 3. In \S 4 we use N-Body simulations to study velocity contrast for
CDM and an HDM like spectrum. \S 5 contains discussion of the new
indicator using spherical model and nonlinear approximations. In \S 6
we compare density and velocity contrast and study the average
relation between them as well as dispersion around it. We also discuss
clustering properties of nonlinear mass and with respect of total
mass.

\section{Evolution of Trajectories}

The problem of gravitational dynamics in an expanding universe can be
simplified considerably for nonrelativistic matter, at scales that
are much smaller than hubble radius. In this domain, we can take
the newtonian limit of relativistic equations. If mass of the universe
is dominated by collisionless matter [ e.g. dark matter] then the
system is entirely characterised by the following equations :
\begin{eqnarray}
\frac{d {\bf u}}{db} & = & - \frac{3}{2} \frac{Q}{b} \left({\bf u} -
{\bf g} \right)\nonumber\\
\nabla^2 \psi & = & \left(\frac{\delta}{b} \right) \nonumber\\
{\bf g} & \equiv & - \nabla \psi \equiv - \frac{2}{3H^2_0 \Omega_0}
\left(\frac{a}{b} \right)  \nabla \varphi\nonumber\\
Q  & = & \left(\frac{\rho_b}{\rho_c}
\right)  \left(\frac{\dot a b}{a \dot b} \right)^2\label{baseqn}
\end{eqnarray}
where ${\bf u} = d {\bf x} / db   $ is the ``velocity''
at time $t$, $\varphi$ is the gravitational potential due to the
perturbed dark matter distribution, ${\bf g}$ is the rescaled
gravitational force, $a(t)$ is the expansion factor, $
\rho_b$  is the background density, $\rho_c$ is the critical density
and $b(t)$ is the growing solution to the equation
\begin{equation}
\ddot b + \frac{2 \dot a}{a} \dot b = 4 \pi G \rho_b b. \label{growth}
\end{equation}
In a matter dominated universe with  $\Omega = 1 $, we have $b=a$
and $Q=1$. [ We shall consider only this case here though
generalisation of our analysis to other models  is
straightforward.]

The equations (\ref{baseqn}) describe a complicated many body system
even in the limit of a smooth gravitational potential. From the
structure of (\ref{baseqn}), we can distinguish four different
epochs in the evolution of clustering --- linear, Zeldovich,
quasilinear and nonlinear.

At sufficiently early time $\delta \ll 1 $ and we can use linear
perturbation theory to study growth of perturbations.  In this limit,
we can easily solve (\ref{baseqn}) and show that
\begin{eqnarray}
\delta \left(a, {\bf x} \right) & = & af \left({\bf x} \right)\nonumber\\
{\bf u} \left(a, {\bf x} \right) & = & u \left({\bf x} \right)  = {\bf g}
\left({\bf x} \right)\nonumber\\
{\bf g} \left({\bf x} \right) & = & - \nabla \psi
\left({\bf x} \right) \nonumber\\
\nabla^2 \psi\left({\bf x}\right) & = & f\left({\bf x}\right) \label{linear}
\end{eqnarray}
Clearly, ${\bf u} \left(a, {\bf x} \right)  $ and ${\bf g} \left(a,
{\bf x} \right) $ are independent of $a$ and $\delta \propto  a $.
Also note that in linear regime, ${\bf u} \left(a,{\bf x}
\right) = {\bf g} \left(a,{\bf x} \right)$. [Velocity
here is, of course, defined in a dynamically
relevant manner as ${\bf u} = \left(d {\bf x} / da \right)$; the
conventional definition of peculiar velocity is ${\bf v} = a \dot {\bf
x} = a \dot a {\bf u} $  and it scales as  ${\rm v} \propto a \dot a
\propto a^{1/2}$. We shall work with ${\bf u}$  since we can always
obtain ${\bf v}$  by a simple rescaling.]

Linear theory becomes invalid when density contrast becomes comparable
to unity, however we can understand some aspects of dynamics by using
the Zeldovich approximation {\cite{za}}. This approximation
extrapolates equality of velocity and force beyond linear regime.
\begin{eqnarray}
\quad {\bf u} \left(a,{\bf q} \right) & = & {\bf g} \left(a_{in},{\bf q}
\right)  = - \nabla \psi \left(a_{in},{\bf q} \right)\nonumber\\
{\bf x} & = & {\bf q} + b\left(a\right) {\bf u}\left(a,{\bf
q}\right)\label{zel}
\end{eqnarray}
Here $\bf q$ is initial position of the particle, also called
its lagrange position. Particles move with constant velocity ${\bf u}$
that is related to the force at its initial position at the initial
epoch. This
approximation compares well with true motion before shell crossing.
Validity of this approximation can be used to infer that ${\bf
u}(a,{\bf x}) \simeq {\bf g}(a,{\bf x})$ in the Zeldovich regime.

Zeldovich regime ends with formation of pancakes and shell crossing.
After shell crossing particles oscillate about pancakes and
form small clumps. These clumps move towards each other or have some
bulk motion towards deep potential wells. Therefore in quasilinear
regime velocities of particles are not aligned with gravitational
force, except in direction of bulk motion of clump to which these
belong.

The mismatch between velocity and force steadily increases till
velocities are randomized inside clumps and dominate over any residual
bulk motion. Such a situation is expected in highly nonlinear and
large clusters that have either virialised or are close to it.

\section{New statistical indicator : Velocity Contrast}

Above discussion shows that mismatch between velocity and
gravitational force is a good indicator of nonlinearity in dynamics
for pancake like models. However, it is as good an indicator for
hierarchical models because nonlinearity in dynamics at small [ mass]
scales does not influence evolution of larger scales.
{\it Shell crossing
is neccessary for formation of nonlinear objects at a given scale,
independent of dynamical state of smaller scales.} Virialised
structures can not form
without shell crossing and mixing in the phase space. Merger of
smaller structures leading to a larger virialised object is always
accompanied by shell crossing at the new scale. Therefore, while
studying a given mass scale we can neglect nonlinearity in dynamics at
much smaller scales. In fact numerical simulations of
hierarchical models are based on the assumption that nonlinearity at
small scales does not influence larger scales.

Mismatch between velocity and gravitational force is a vector quantity
and is therefore difficult to handle and interpret. In addition it is a
dimensional quantity and numerical value of the mismatch must be
compared with something else for making a model independent estimate
of the level of nonlinearity in dynamics. These considerations lead us
to suggest the following form for velocity contrast :
\begin{equation}
D_{gu} \equiv {\left({\bf u} - {\bf g} \right)^2 \over {\rm u}^2 }
\label{velcont}
\end{equation}
In this equation we can think of ${\bf u}$ and ${\bf g}$ as velocity
of a given particle and force acting on it, respectively.
Alternatively if smooth velocity and force fields are given,
(\ref{velcont}) defines a scalar field $D_{gu} \left(a, {\bf x}
\right)$. These two descriptions are equivalent for our purpose. If
linear theory [ or Zeldovich approximation] is valid for most
particles [ regions], then this quantity will be nearly zero for a
large fraction of particles [ mass]. With further evolution of
clustering, more and more particles [ mass] acquire significant values
for this parameter. Regions containing particles [ mass] with
$D_{gu}$  larger than a threshold will exhibit highly nonlinear
dynamics and a study of these regions will offer insight into the study of
nonlinear clustering. Luminuous objects that we see have nonlinear
density contrasts, therefore it is more meaningful to compare
clustering properties of nonlinear objects in simulations with
observations. $D_{gu}$ provides a simple method for selecting
nonlinear structures in a simulation.

This algorithm has two specific advantages over the method of density
threshold or the friends of friends algorithm where one finds
nonlinear structures by requiring local density to be higher than some
cutoff or particles to be within some linking length of its nearest
neightbour. Density threshold [ linking length] for an
object that has decoupled from expansion depends on the local symmetry
of collapse, whereas deviations of
velocity from accleration depend only on the stage of dynamical
evolution. When particles in a particular region are counted,
in order to ascribe a density to that region, one  does  not take into
account velocities of particles. A fast moving particle which is
merely passing through a region will be counted in this process {\it
even if it is following a Zeldovich trajectory with } ${\bf u} = {\bf
g} $.  In using  $D_{gu}$  as an indicator we exclude such particles
until the alignment between ${\bf u}$  and ${\bf g}$ is disrupted. Our
algorithm will select ``nonlinear particles'' in underdense regions as
well. These are found in regions where pancakes are forming in
otherwise underdense regions or where density is low but shear
is important.

Velocity contrast suggests another method for studying approximation
schemes. Note that $D_{gu}$ directly
characterises what could arguably be considered the most significant
factor in formation of bound structures: {\it the ability of a
local mass inhomogeniety to pull back particles towards it and therby
increase the local potential depth.} It is precisely the failure to do
this which makes Zeldovich approximation break down at the onset of
significant nonlinearity. For comparison, consider the ``frozen
potential approximation'' [ FPA hereafter; see Bagla and Padmanabhan
\shortcite{fpa}; also suggested independently by Brainerd, Scherrer
and Villumsen \shortcite{lep}] which may be thought of as a logical
continuation of Zeldovich approximation: in ZA, velocities are frozen
to initial values and force is ignored; in FPA, gravitational force is
frozen to the initial value and the particles are moved in this given
background potential. Studies have shown that FPA correctly reproduces
the behaviour of particles near the mass concentrations since
$\left({\bf u} - {\bf g} \right)$ is not pre-assigned to vanish  in
this approach. Therefore the structure of pancakes in FPA is
similar to that in N-body simulations; pancakes do not thicken though
they are not as thin as those seen in N-body simulations. Comparison
of $D_{gu}$ in FPA and N-body will be useful in understanding
both.

Note that this statistical indicator takes into account both
the velocity and force. It is possible to devise approximation
schemes, like frozen flow {\cite{ffa}}, adhesion model {\cite{adhes}},
etc. which will move particles to the right regions but will give a
physically unacceptable picture for the velocity field. In our
opinion, approximation schemes should also provide correct velocity
field if they have to offer insight into dynamics. $D_{gu}$ may be
used to discriminate between the dynamical content of different
approximation schemes.

Velocity contrast can be used to compare analytical models with N-body.
For simple analytic models of structure formation like spherical top
hat it is possible to compute $D_{gu}$ as a function of the density
contrast $\delta$. As we shall see this helps one in forming an
intutive picture of the nonlinear evolution.

\section{N-body Simulations}

To understand velocity contrast, we study it using N-Body
simulations. Use of analytical methods for this study is made
difficult by the fact that velocity contrast vanishes in linear
regime. In this section we present results for two representative
models of structure formation.

First model we consider is standard
unbiased CDM normalized to COBE. Simulation of this model used a
particle-mesh code on a $128^3$ box with $128^3$ particles. The
size of the box in physical units is $90 h^{-1} Mpc$. Figure 1 shows
projected density field and velocity contrast field for slices $14
h^{-1} Mpc$ thick. These slices corresponds to reshifts $Z=3$, $Z=1$
and $Z=0$.

\begin{figure}
\epsfxsize=6truein\epsfbox[48 26 564 765]{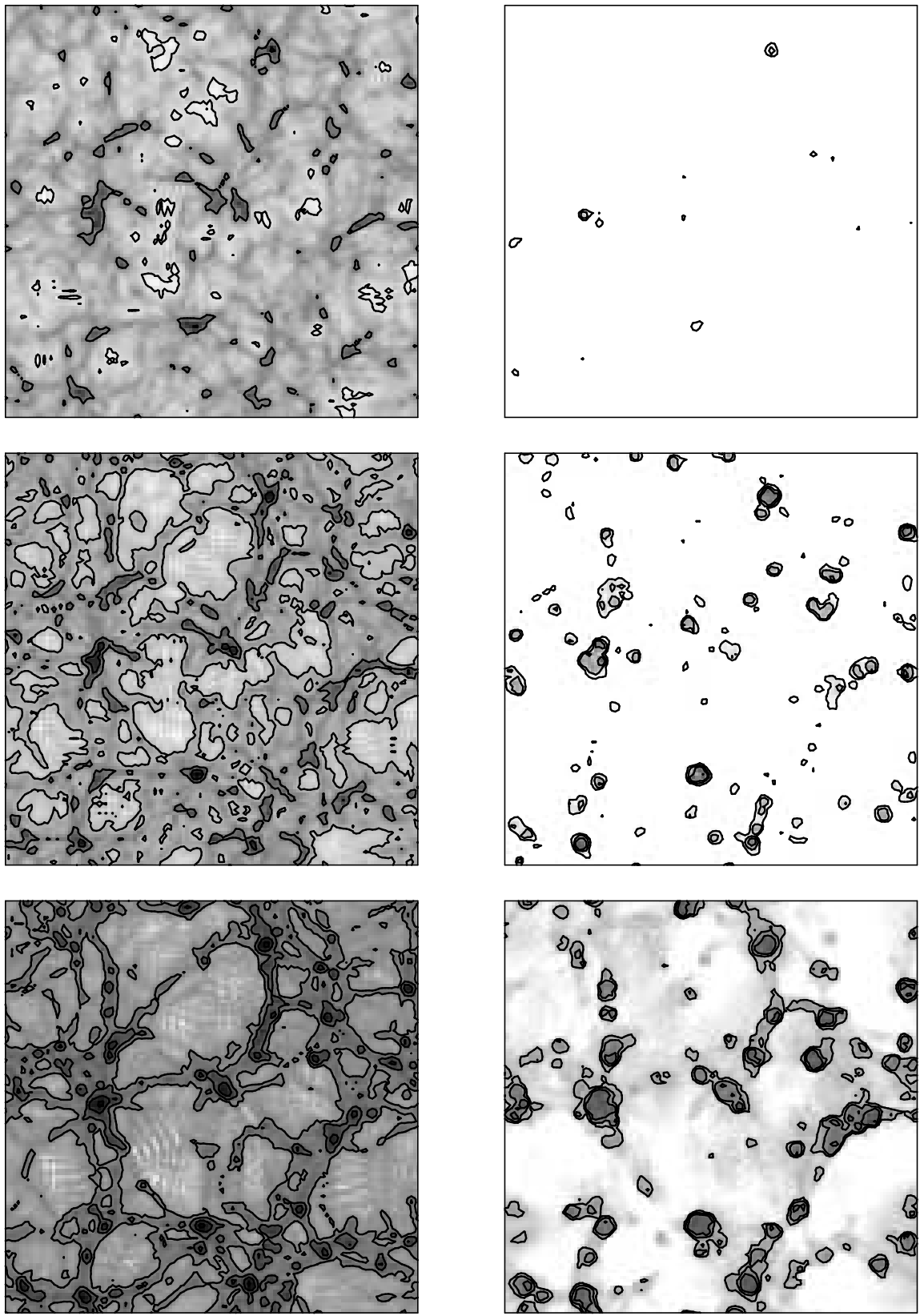}
\caption{These figures show grey scale maps and contours of projected
density and velocity contrast for slices taken from N-Body
simulations of CDM. Left frames show the density field and right
frames show velocity contrast field for the corresponding slices. Top
frames are for $Z=3$, middle frames are for $Z=1$ and the bottom
frames show the same slice at $Z=0$.}
\end{figure}

Comparison of panels at a given redshift shows that density and
velocity contrast select almost the
same set of nonlinear regions. These tend to
disagree for some regions that appear dense to the eye but have not
reached a sufficiently high level of nonlinearity in dynamics. On the
other hand, some regions in what appears to be a void have a high
value of velocity contrast. Highly nonlinear regions appear to have a
larger radius of influence as diagnosed by velocity contrast, in
comparison with density. This may indicate that velocity
contrast does not increase substantially after virialisation, so that
almost all members of a virialised cluster have similar values of
velocity contrast even though density varies rapidly with distance
from the centre. Evolution with redshift follows the expected pattern
with more regions becoming nonlinear at later times.

Regions selected for contours of high velocity contrast tend to be
spherical. This indicates that planar and filamentary structures can
not attain a very high level of nonlinearity in dynamics --- for
nonspherical structures there is always some direction in which bulk
velocity dominates over random motions.

An interesting feature that emerges from figure 1 is that velocity
contrast filters out voids [ apart from regions within them where
large shear leads to formation of structures] more effectively as
compared to density contrast.

\begin{figure}
\epsfxsize=6truein\epsfbox[48 26 564 765]{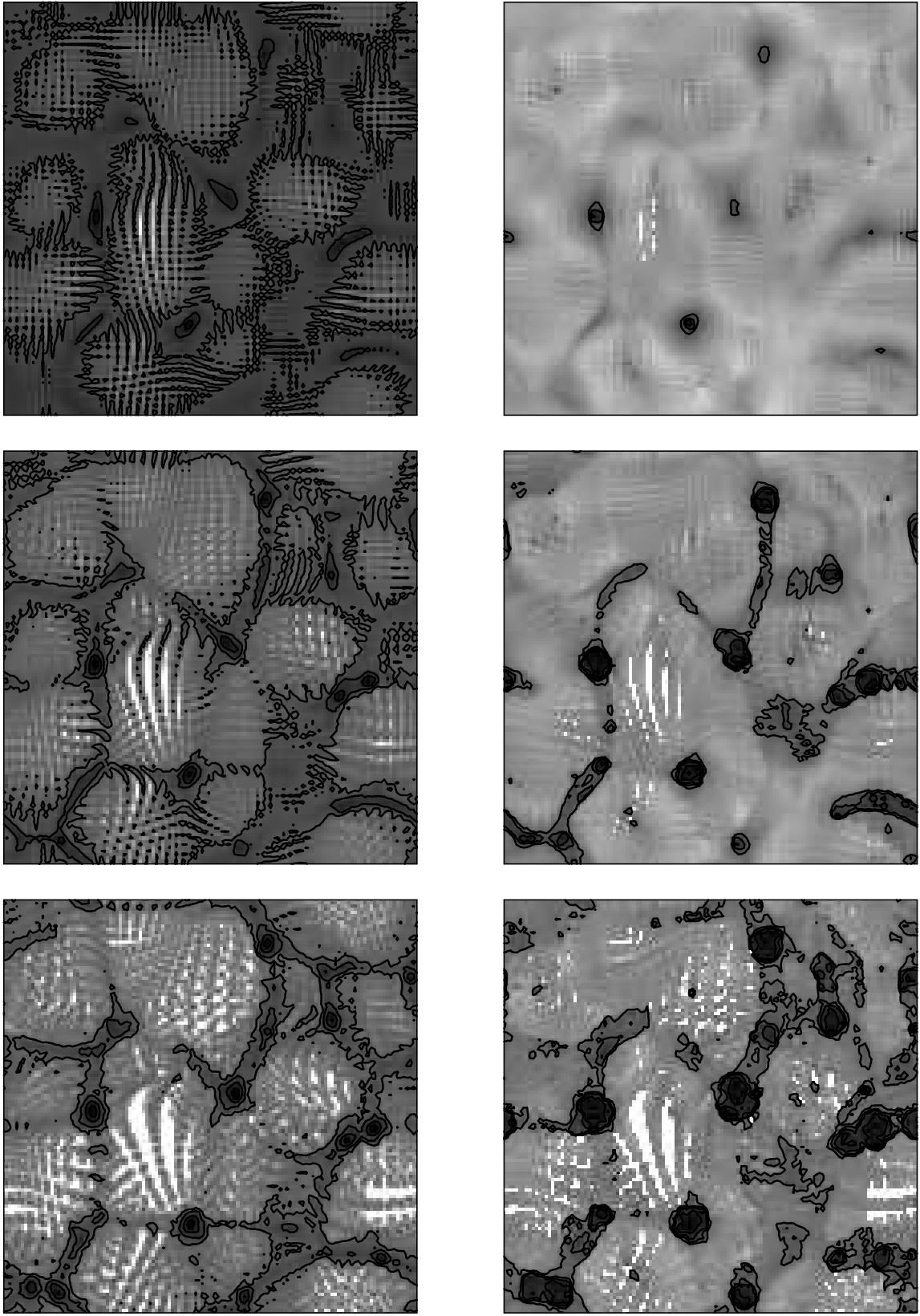}
\caption{Same as figure 1 but for model with a gaussian power
spectrum.}
\end{figure}

The CDM model is a prime example of hierarchical clustering scenario
with power at all scales and effective index varying from $-3$ to
$1$. To see how this algorithm works in other cases, we studied the
other extreme example. We considered a hypothetical universe which has
power peaked at one dominant scale. The power spectrum for this model
is taken to be
\begin{equation}
P(k)= \frac{A}{\Delta k (2\pi)^{1/2}}
\exp\left[-\frac{\left(k-k_0\right)^2}{2(\Delta k)^2}\right]\label{pofk}
\end{equation}
with $k_0 = \pi/16$ and $\Delta k = \pi/64$. Clearly, this has power
only in a band of width $\Delta k$ peaked at $k_0$. We chose
$A=68000$ so that the linearly extrapolated amplitude of the peak is
unity at $Z\sim2$, giving us sufficient coverage of both the linear
and nonlinear regime. Our choice of $k_0$ and $\Delta k$ ensures that
the power
is peaked at a large scale and it is concentrated in a very narrow
range of scales. Note that this spectrum is somewhat similar to HDM
models as far as the peak and smaller scales are concerned; standard
HDM has $P \left(k \right)  \propto k $ for small $k$ but this model
has an exponentially low power at  small $k$. This is not
very important since a $k^4$ spectrum will be generated due to
dynamics {\cite{k4tail}}. The evolution of this model provides an
interesting paradigm for structure formation and will be discussed
in detail elsewhere \cite{onemode}. Here we merely use this model to
study velocity contrast so as to test it in the widest possible
range. This simulation used a $128^3$ box with $64^3$ particles.

Figure 2 shows projected density and velocity contrast fields for
this model. The thickness of the slice used here is $20L$, where $L$ is
one grid length in the simulation box. Panels of this figure follow
the pattern of panels in figure 1, showing that the general behaviour
of velocity contrast is
independent of the type of model considered. Particular values of
velocity contrast for particles will of course depend on the mass
resolution used in simulations, but the smoothed field has same
properties for a large range in mass scales : $M_{particle}\ll M \ll
M_{box}$.

\begin{figure}
\epsfxsize=6.5 true in\epsfbox[48 279 564 765]{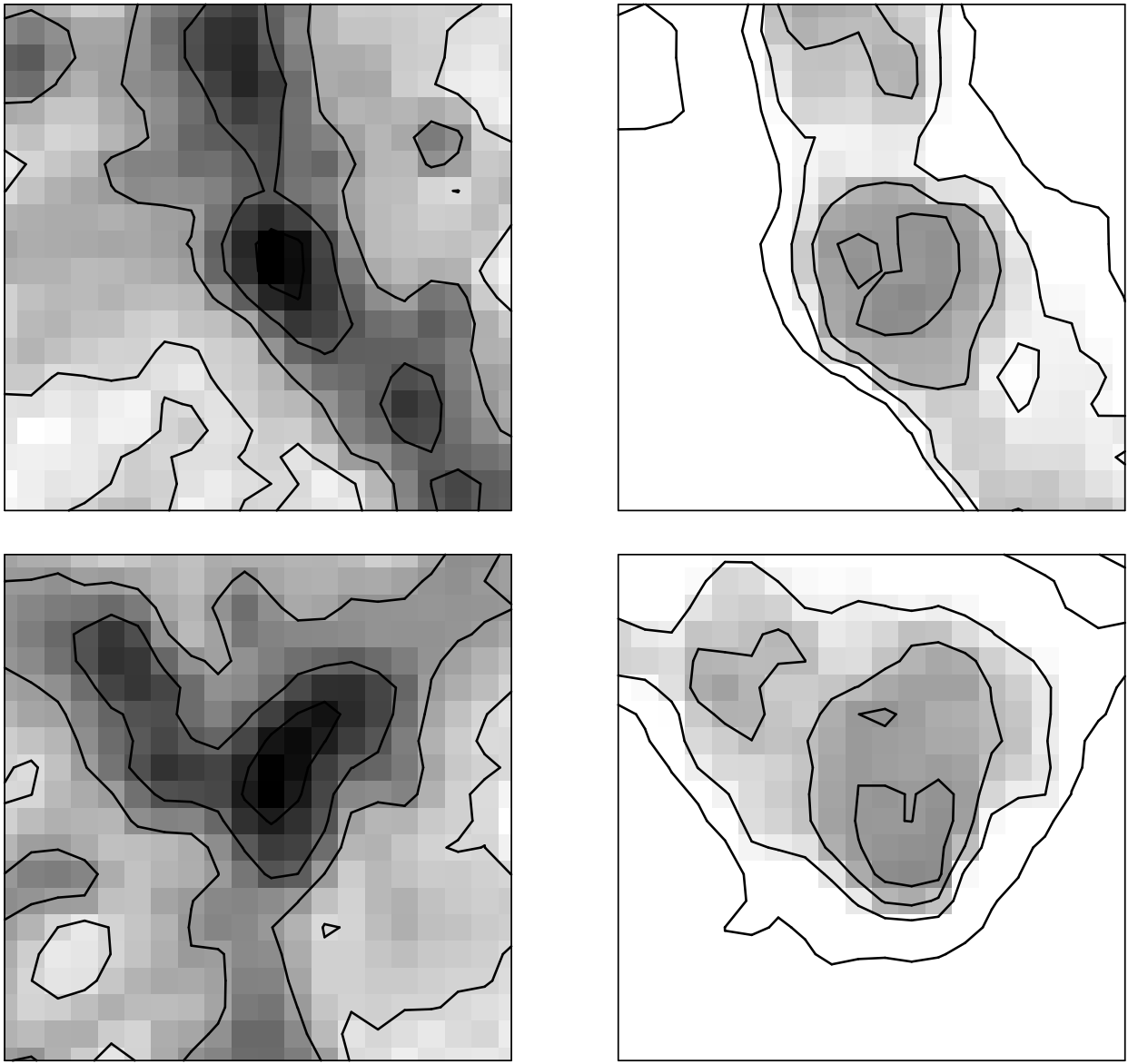}
\caption{These figures show close up of two clusters for the CDM
simulation. Left panels show projected density field and right panels
show projected velocity contrast field. A cube of size $14h^{-1}Mpc$
centered at the clusters was used for these plots.}
\end{figure}

One key difference between density and velocity contrast is that
density is a function of space and is same for particles located at
same position whereas velocity contrast can be different for such
particles. This feature of velocity contrast suggests that it can give
us insight about some facets of clustering in highly nonlinear regions that
are inaccessible through density contrast. We bring out these
differences by studying some clusters in greater detail. Figure 3
shows close up of two rich clusters from the CDM simulation. For each
cluster we have plotted projected density and velocity contrast fields
for a cube of size $14h^{-1}Mpc$ centered at the cluster. Comparison
of panels for density and velocity contrast shows that for high
density regions information content of these indicators differs
significantly. Contours of equal density indicate positions of density
peaks and outlines of pancakes, whereas contours of velocity contrast
show us regions that have significantly nonlinear dynamics.

Some regions that appear as pancakes or small density peaks are not
seen in contours of
velocity contrast, particles in these regions have not undergone any
shell crossing but are only falling into the cluster. In all such
regions, bulk flow towards the nearby cluster dominates over any random
motions within the infalling mass. Many density
peaks show complex substructure within. This can result from two
scenarios : either there has been some recent merger of two small
clusters and these have not mixed sufficiently or a group of particles
with a much lower level of nonlinearity in dynamics has fallen into the
cluster and it has not become an integral part of the cluster in
dynamical sense.

Figure 4 shows a similar close up of a cluster for the model with
gaussian power spectrum. Here, the
substructure is not as complex as in figure 3 as there was no small scale
power in the initial spectrum. However, broad features are similar for
both these models.

\begin{figure}
\epsfxsize=6.5 true in\epsfbox[48 530 564 765]{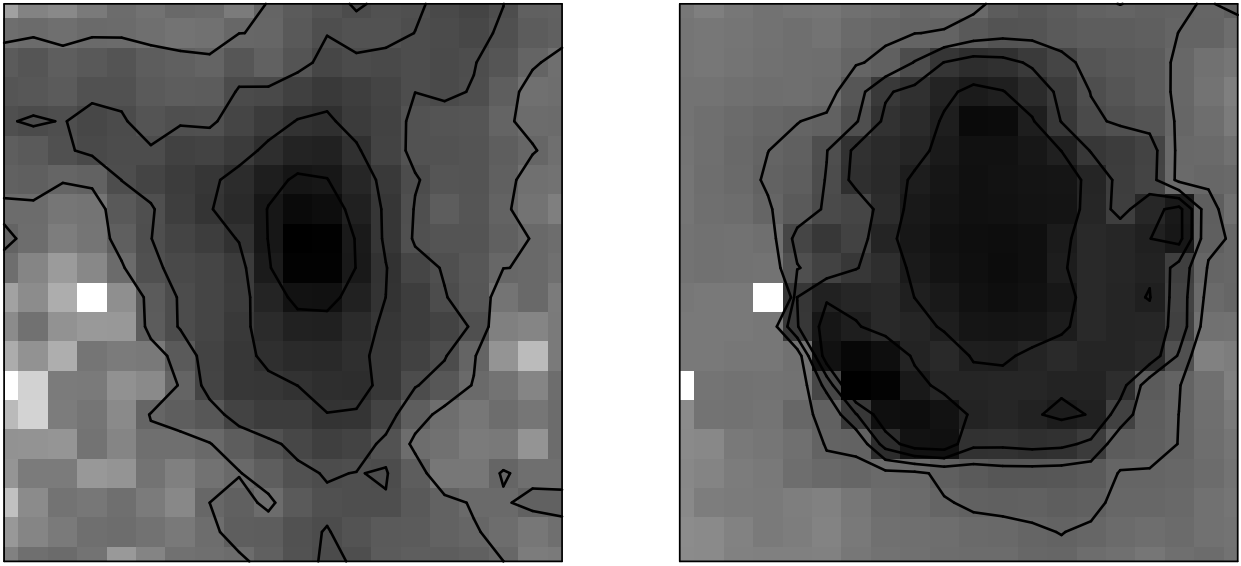}
\caption{Same as figure 3 but for model with gaussian power spectrum.
We have shown close up of only one cluster for this model.}
\end{figure}

We now give the above results in a less picturesque manner. Figure 5
shows the fraction of particles with velocity contrast above a threshold as a
function of threshold velocity contrast at different
epochs for the CDM model. Same figure shows corresponding curves for
the second model as dashed lines. For reference, we have labelled the
equivalant density contrast for top hat spherical collapse on the top
axis. These curves show that as clustering proceeds the fraction of
particles with $D_{gu}$ larger
than a given threshold increases. Also, the fraction of particles selected
with a high cutoff are significantly less than those selected by a
smaller value of $D_{gu}$. These features are expected in any
indicator of nonlinearity and it is heartening to see that velocity
contrast satisfies these criterion. These features can also be seen,
at a qualitative level, in figures 1 and 2.

\begin{figure}
\epsfxsize=6.5 true in\epsfbox[85 400 510 752]{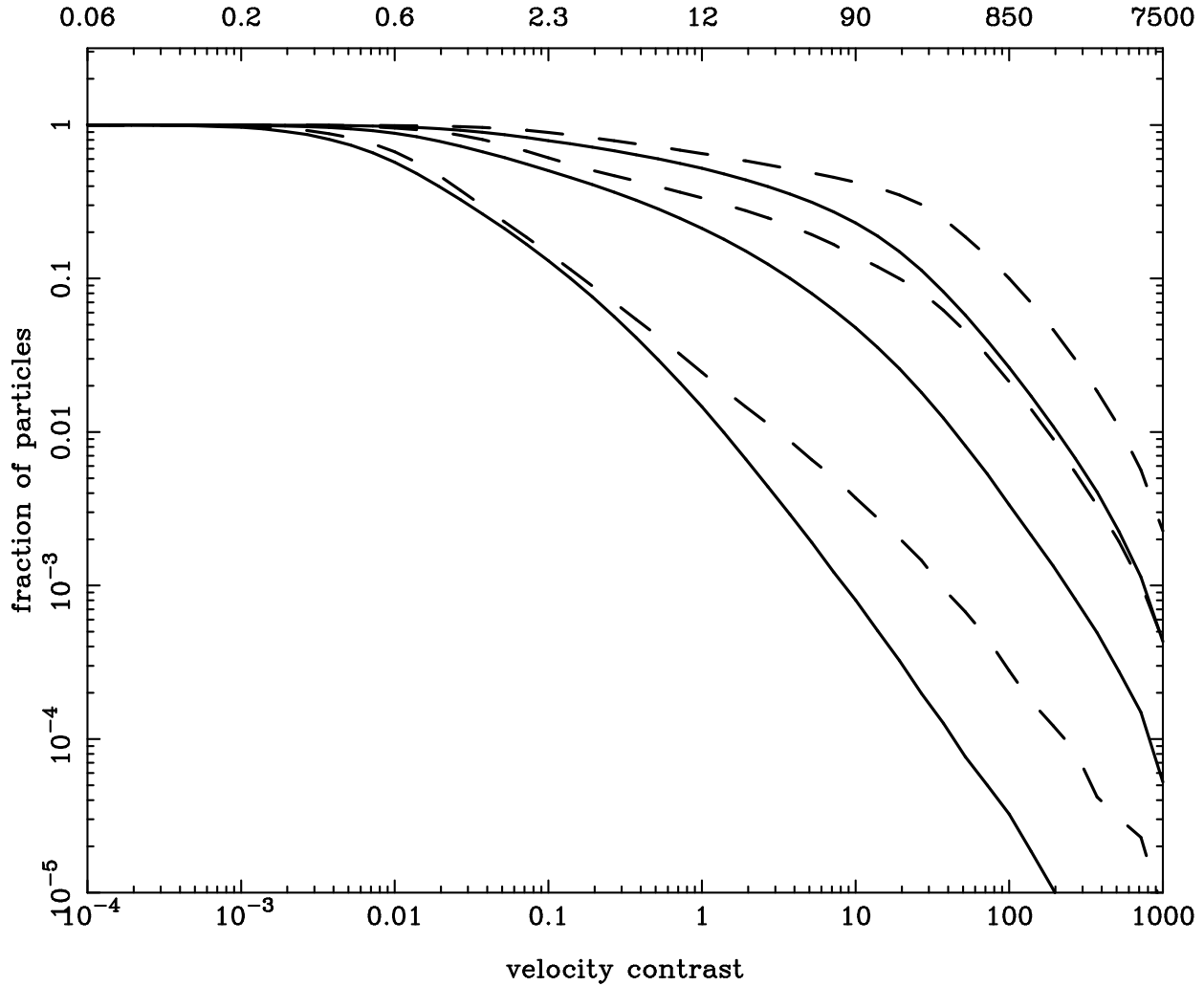}
\caption{Fraction of particles above a threshold velocity contrast are
shown as a function of threshold for three redshifts : $Z=3,\, 1$ and $0$.
Thick lines are
for CDM and dashed lines are for model with gaussian power
spectrum. Top axis has been labelled by the equivalant density
contrast for spherical top hat collapse.}
\end{figure}

\section{Analytical Models and Nonlinear Approximations}

In this section we study velocity contrast with help of nonlinear
models and analytical approximations. Spherical top hat collapse is
one model of nonlinear collapse where it is possible to calculate
almost all quantities of interest. Here we compute velocity contrast
as a function of density contrast for this model. Initial velocities
are related to initial density contrast through the potential as
\begin{eqnarray}
{\bf u}\left( a, {\bf x}\right) & = & {\bf g}\left( a, {\bf
x}\right) = -\nabla\psi\left( a, {\bf x} \right) \nonumber\\
\nabla^2 \psi\left( a, {\bf x} \right) & = & \frac{\delta\left( a
\right) }{a}\label{sthinvel}
\end{eqnarray}
Here {\bf x} is the comoving coordinate and relates to the proper
coordinate {\bf r} in the usual way. These initial conditions put the
system into growing mode and this leads to the following solution.
\begin{eqnarray}
r & = & \frac{r_{max}}{2} \left(1-\cos\theta\right)\nonumber\\
a & = & \frac{3 a_i}{5 \delta_i}\left[ \frac{3}{4} \left( 1 + \delta_i
\right) \left( \theta - \sin\theta \right) \right]^{2/3}\nonumber\\
\delta & = & {\frac{9}{2}} {\frac{\left(\theta -
\sin\theta\right)^2}{\left(1-\cos\theta\right)^3}} - 1\label{sthdel}
\end{eqnarray}
We can also write down expressions for ${\bf g}$ and ${\bf u}$,
leading to an expression for velocity contrast.
\begin{eqnarray}
{\bf g} & = & - \frac{\delta}{3 a^2} {\bf r}\nonumber\\
{\bf u} & = & - \frac{1}{3a} \frac{\partial \ln(1+\delta)}{\partial a}
{\bf r}\label{sthgu}
\end{eqnarray}
{}From these, given a value of velocity contrast we can determine
the corresponding phase angle  and solve for density
contrast. In figure 6 we have plotted density contrast as a function
of velocity contrast for this model.

\begin{figure}
\epsfxsize=6.5 true in\epsfbox[35 397 558 748]{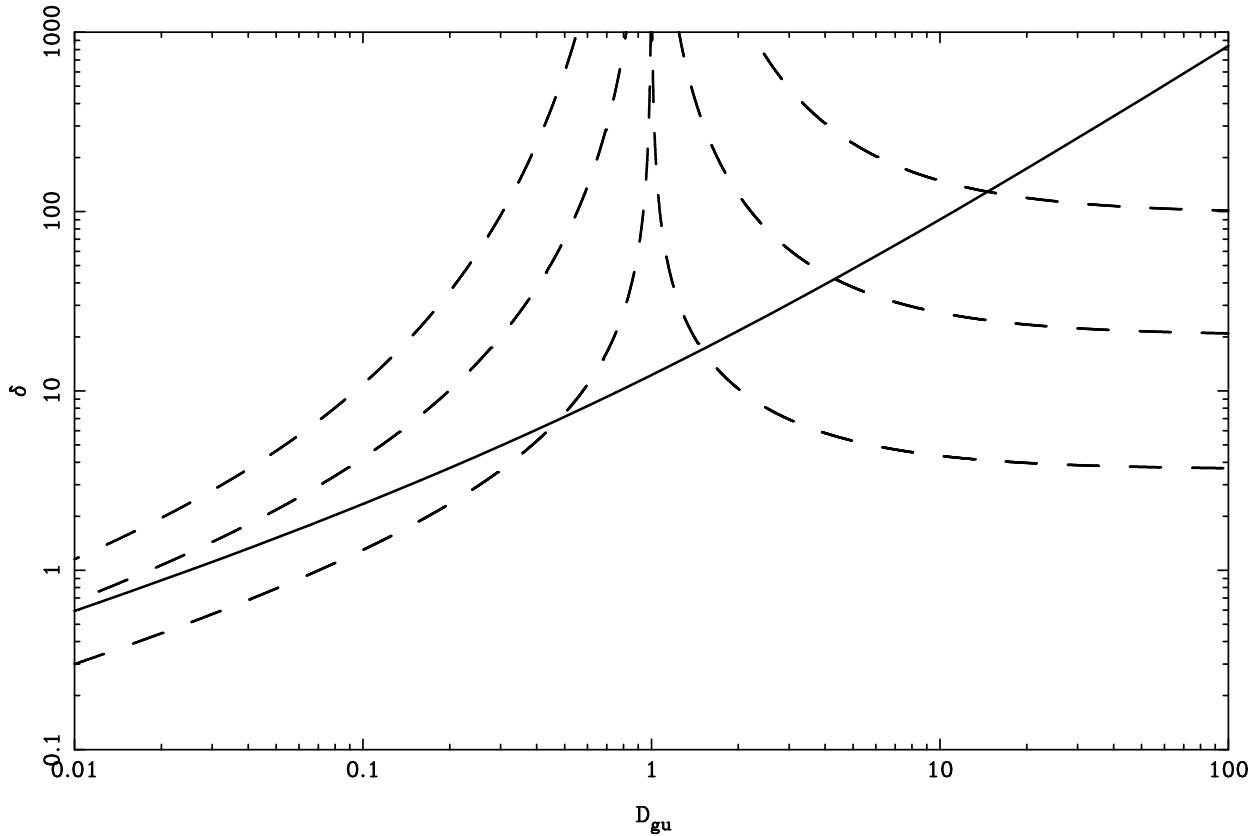}
\caption{Thick line shows density contrast as a function of velocity
contrast for spherical top hat collapse. Dashed lines show curves for
spherical, cylindrical and planar collapse in the frozen potential
approximation. The lowest curve corresponds to planar collapse and the
highest to spherical collapse.}
\end{figure}

One certainly does not expect STH to be a good model for generic nonlinear
collapse and it is worthwhile to consider evolution of velocity
contrast for symmetric collapse in other approximations. In figure 6
we have also plotted curves computed using FPA for one , two
and three dimensional models; that is, in each case we solve for
particle trajectory around the potential produced by a sheet [1-D
motion, sheet is 2D], filament [2D motion, source is 1D] or a
spherical region of constant density. These curves show that for a
given velocity contrast, density is lowest for planar collapse and
highest for spherical collapse. The ratio of densities for $D_{gu{\bf
c}} = .1$ is $1:3.2:8.9$. This clearly shows that velocity contrast is
a better algorithm for selecting nonlinear regions as it defines
nonlinearity as deviations from trajectories in the linear regime. It
is clear that a density threshold for selecting nonlinear structures
will either select spherical objects that have not turned around or
shell crossed, or it will miss out mildly nonlinear planar
structures. Another feature that is seen in these graphs is that after
shell crossing, density decreases while velocity contrast
increases. When the particle turns back towards the pancake, it is
possible for it to return to very small values of $D_{gu}$ but this
happens only because in FPA we are using the initial gravitational
force. In exact evolution, ${\bf g}$ evolves sufficiently
to rule out small values of $D_{gu}$. Another feature that is seen
here is that shell crossing occurs in FPA at $D_{gu} = 1$ as
acceleration vanishes at the point where pancake is formed. [In FPA we
use initial acceleration, which vanishes at the caustic in symmetric
cases. In a general case, there will be some residual acceleration
along the caustic.]

Velocity contrast can be studied within the framework of adhesion
model {\cite{adhes}}. In this model the equation of motion is
modified to include an ad hoc viscous term that ensures thin pancakes.
The equation of motion is
\begin{equation}
\frac{\partial {\bf u}}{\partial a} + \left({\bf
u}.\nabla\right){\bf u} = \nu \nabla^2 {\bf u}\label{adhesion}
\end{equation}
A comparison with (\ref{baseqn}) shows that viscosity term tries to mimic
gravitational force, and
\begin{equation}
{\bf g}_{eff} = {\bf u} + \frac{2b}{3Q}\nu\nabla^2 {\bf u}\label{geff}
\end{equation}
In this equation, gravitational force is some kind of an effective
acceleration modelled by the viscosity term.

The general solution to (\ref{adhesion}) is well known and it allows us to
compute ${\bf u}$ for any particle, given the initial velocity
potential. We can use this solution to compute velocity contrast, for the
effective force in adhesion model. It is apparent that $D_{gu}$
is significantly large only at the caustics in the limit $\nu=0$. At
present we are studying $D_{gu}$ as a function of $\nu$, which could
lead to  an understanding of the origin of effective viscosity in the
adhesion model.

\section{Evolution and Clustering of Nonlinear Structures}

Velocity contrast and its relation with density contrast is shown in
figure 7 where we have plotted contours of equal population on the $(1
+ \delta) - D_{gu}$ plane. This plot shows contours for two epochs
each for the two models we are using in this paper : CDM and gaussian
power spectrum. For comparison we have also plotted the curve obtained
from STH on each of the panels. These plots show that there is an
average relation between velocity contrast and density that is
independent of epoch and model, and holds for a large range of
scales. For CDM simulations, it is apparent that the relation evolves
at high velocity contrast and the density for a given velocity
contrast increases with time. This could imply a shift from mostly
planar collapse occuring at early epochs towards more instances of
spherical collapse at later stages.

There is a large dispersion about the average relation between
density and velocity contrast. Part of this dispersion arises from the
fact that collapse of various kinds, from planar to spherical, is
occuring in the simulation volume. As indicated by relation of
velocity contrast with density for symmetric collapse in FPA, the
variation in density contrast for a given velocity contrast may be as
large as an order of magnitude. Contours for the second model show
a very large dispersion at large velocity contrast. This indicates
ongoing collapse of objects with different local symmetries.

\begin{figure}
\epsfxsize=6.5true in \epsfbox[45 270 556 750]{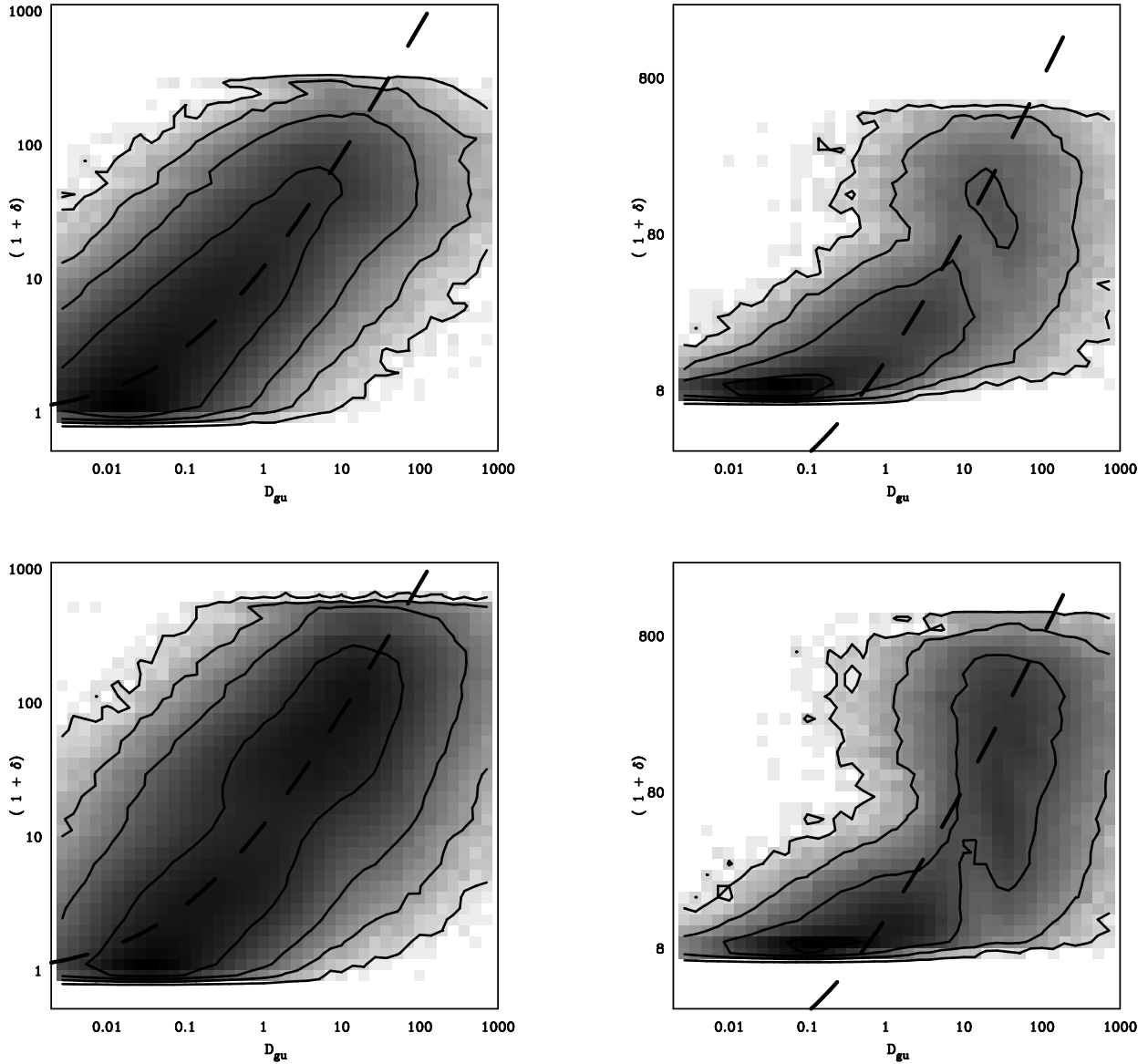}
\caption{Contours of equal population on the density -- velocity
contrast plane. Left panels show these contours for CDM and right
panels are for model with gaussian power spectrum. Top panels are for
$Z=1$ and bottom panel is for $Z=0$. Dashed line in each panel shows
relation for STH collapse.}
\end{figure}

Velocity contrast is a good indicator of nonlinearity. In a simulation
output we can identify nonlinear regions by requiring velocity
contrast to be higher than some threshold. Regions selected in this
manner will in general have clustering properties that differ from
those of the total underlying mass. We study clustering properties of
these regions and define a scale dependent bias parameter. Only those
regions where particles have undergone shell crossing [ at the scale
of a galaxy] can we form virialised structures like galaxies that we
see. Therefore, the clustering properties of regions with velocity
contrast above a threshold are essentialy clustering properties of
regions that can host visible structures. Figure 8 shows averaged
correlation function for the underlying mass and also for regions with
velocity contrast above a threshold in a CDM simulation at
Z=0. Threshold values used here were $0.1$, $1$ and $17$. The averaged
correlation function $\bar\xi$ and bias $b$ are defined here as :
\begin{eqnarray}
\bar \xi \left(a, x \right)= \frac{ 3}{ x^3 } \int\limits^x_0
\xi \left(a, y \right)y^2 d y\nonumber\\
b^2 \left( x\right) = \frac{\bar\xi\left(x ; D_{\bf gu} > D_{{\bf
gu}c}\right)}{\bar\xi\left(x ; {\rm all\,\,  particles}\right)}\label{bias}
\end{eqnarray}
where $\xi$ is the two point correlation function. The relative bias
decreases with the scale, approaching unity at very large scales, and
increases with the cutoff used in selecting particles. This scale
dependent bias may be used in understanding relative bias between
different kinds of objects, like different populations of galaxies and clusters
of galaxies. Unlike the method of peaks in the initial distribution,
the algorithm described in this paper clearly picks out regions that
are dynamically ``nonlinear'' and therefore are potential sites for
galaxy formation. It may be possible to model the environmental
effects in galaxy formation as $D_{gu}$ describes the level of
``churning'' for nonlinear regions which is likely to be an important
factor.

\begin{figure}
\epsfxsize=6.5 true in\epsfbox[30 524 563 752]{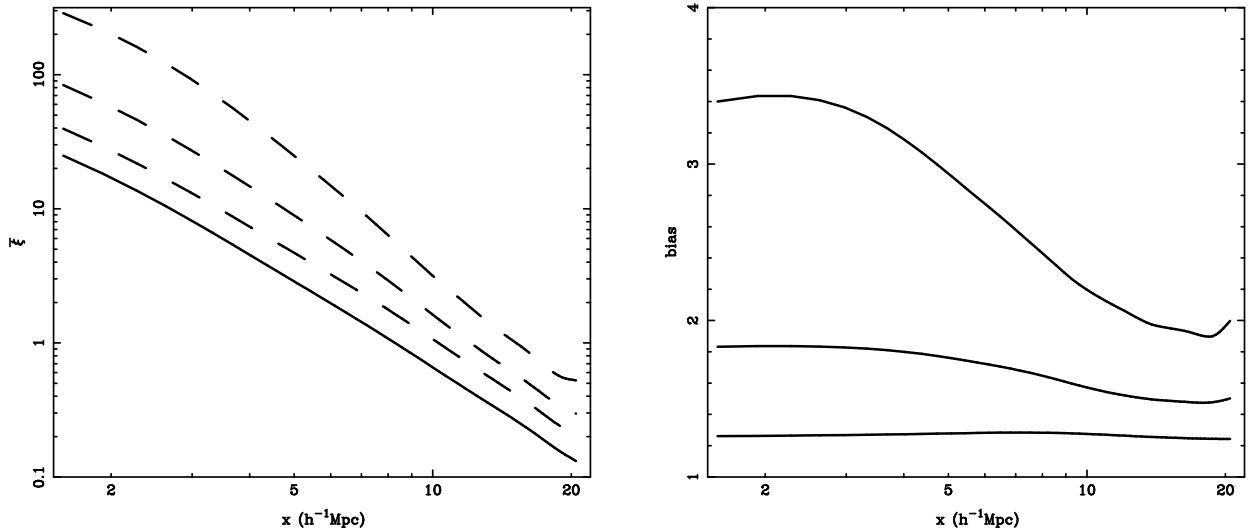}
\caption{$\bar\xi$ vs. $x$ for CDM at $Z=0$. We have also plotted
$\bar\xi$ for particles with $D_{\bf gu} > .1,\, 1$ and $17$. Left
panel shows bias parameter for these threshold values as a function of
scale.}
\end{figure}

\section{Conclusions}

The purpose of this note was to introduce this new statistical
parameter, study its behaviour and establish a prima facie case that
it is worth being considered further.  It will be interesting to see
whether one can develop an analytic model for the evolution of this
statistical indicator -- or a closely related one. This and related
questions are under investigation.

We are also studying the bias parameter introduced here in detail,
including effect of mass resolution etc. This work is in progress and
will be reported elsewhere {\cite{bias_pr}}.

JSB is being supported by the Senior Research Fellowship of CSIR India.

\label{lastpage}

\end{document}